\RequirePackage[2020-02-02]{latexrelease}
\documentclass[twocolumn,prb,amsmath,amssymb,superscriptaddress]{revtex4}
\usepackage{graphicx}
\usepackage{bm}
\usepackage{array}

\begin{document}

\title{Generalized method of image dyons for quasi-two dimensional slabs with ordinary - topological insulator interfaces}

\author{Jose L. Movilla}
\email{movilla@uji.es}
\affiliation{Dept. d'Educaci\'o i Did\`actiques Espec\'ifiques, Universitat Jaume I, 12080, Castell\'o, Spain}

\author{Juan I. Climente}
\affiliation{Dept. de Qu\'imica F\'isica i Anal\'itica, Universitat Jaume I, 12080, Castell\'o, Spain}

\author{Josep Planelles}
\affiliation{Dept. de Qu\'imica F\'isica i Anal\'itica, Universitat Jaume I, 12080, Castell\'o, Spain}

\date{\today}

\begin{abstract}

Electrostatic charges near the interface bewteen topological (TI) and ordinary (OI) insulators induce magnetic fields in the medium that can be described through the so-called method of image dyons (electric charge - magnetic monopole pairs), the magnetoelectric extension of the method of image charges in classical electrostatics. Here, we provide the expressions for the image dyons and ensuing magnetoelectric potentials in a system comprised by two planar-parallel OI-TI interfaces conforming a finite-width slab.
The obtained formulae extend earlier work in that they account for all different combinations of materials forming the slab and its surroundings, including asymmetric systems, as well as all possible combinations of external magnetization orientations on the interfaces.
The equations are susceptible of implementation in simple computational codes, to be solved recurrently, in order to model magnetoelectric fields in topological quantum wells, thin films, or layers of two-dimensional materials. We exemplify this by calculating the magnetic fields induced by a point charge in nanometer-thick quantum wells, by means of a Mathematica code made available in repositories.

\end{abstract}

\maketitle

\section{Introduction}

Since first theoretically predicted\cite{kane05,bernevig06,fu07} and experimentaly discovered,\cite{konig07,hsieh08} topological insulators (TIs) have attracted increasing attention in condensed matter physics because of their novel, exotic properties, as well as of their potential applications \cite{tian17}. TIs resemble ordinary insulators (OIs) in the sense that they present bulk bandgaps, but differ in that TIs also present gapless, metal surfaces with boundary (edge or surface) states arranged in a massless Dirac spectrum. 
Noticeably, this spectrum stands out for being extremely robust against perturbations and/or surface conditions as long as time-reversal (TR) symmetry is preserved, for this robustness comes from bulk properties.
When TR symmetry is broken -say by means of a magnetic perturbation (applied field and/or film coating)- in the vicinity of the interfaces, a gap opens in the otherwise topologically protected spectrum of the TI surface, leading to unusual electromagnetic and magnetotransport effects.\cite{Qi08} One such effect is the so-called topological magnetoelectric effect (TME), a collection of phenomena where magnetic fields become the source of electric fields and viceversa. The electromagnetic response is then described by modified Maxwell equations, which include additional terms that couple an electric field to a magnetization and a magnetic field to a polarization of the medium. Formally, such a description can preserve the usual form of the Maxwell equations, but with modified constitutive relations which contain additional cross-terms.

Thus, similarly to ordinary insulators, the Maxwell equations for TI can be written (in differential form and Gaussian atomic units) as

\begin{eqnarray}
\label{eqGauss}
\nabla \cdot {\bf D} &=& 4 \pi \rho,\\ 
\label{eqFaraday}
\nabla \times {\bf E} &=& -\frac{1}{c} \frac{\partial {\bf B}}{\partial t},\\
\label{eqGaussMagn}
\nabla \cdot {\bf B} &=& 0,\\
\label{eqAmpere}
\nabla \times {\bf H} &=& \frac{1}{c} \frac{\partial {\bf D}}{\partial t} + \frac{4 \pi}{c} {\bf J}.
\end{eqnarray}

However, when topological insulators are involved, the electric displacement vector ${\bf D}$ is modified by the magnetic induction ${\bf B}$ and the magnetic field intensity ${\bf H}$ is in turn influenced by the electric field. Then, the constitutive relations ${\bf D} = \epsilon {\bf E}$ and ${\bf H} = {\bf B}/\mu$ must be modified according to \cite{Qi08,Qi09,Qi11,MartinRuiz,Campos}

\begin{eqnarray}
\label{eqConstD}
{\bf D} &=& \epsilon {\bf E} - P \frac{\theta \alpha}{\pi} {\bf B}\\ 
\label{eqConstH}
{\bf H} &=& \frac{{\bf B}}{\mu} + P \frac{\theta \alpha}{\pi} {\bf E} 
\end{eqnarray}

\noindent where $\alpha = e^2/(\hbar c) \approx 1/137$ is the fine-structure constant, $\epsilon$ and $\mu$ the relative dielectric constant and magnetic permeability, and $\theta$ the magnetoelectric polarizability. In bulk TI, $\theta=\pi$, while in ordinary media $\theta=0$, thus recovering the usual constitutive relations. Finally, $P=\text{sgn} [{\mathbf M} \cdot {\mathbf n}]$ determines the sign of the magnetoelectric coupling. Here, ${\mathbf M}$ is a (external) surface magnetization breaking locally the TR symmetry, and ${\mathbf n}$ the suface unit vector pointing out of the TI.\cite{Qi09,Campos,sekine}

The unusual physics of the TME has sparked several studies over the last years aiming at gaining intuition on how these fields behave in different systems with different geometries. \cite{bernevig06,Qi09,Qi11,Campos,jp22,Yng-zulicke,fechner14,fechner19,MartinRuiz16,Nogueira2022} Whenever planar interfaces are involved, the method of choice to model the TME is the so-called method of image dyons, the axionic counterpart of the well-known method of image charges in electrostatics.\cite{jackson}

\begin{figure}[htb]
\begin{center}
\resizebox{0.8\columnwidth}{!}{\includegraphics{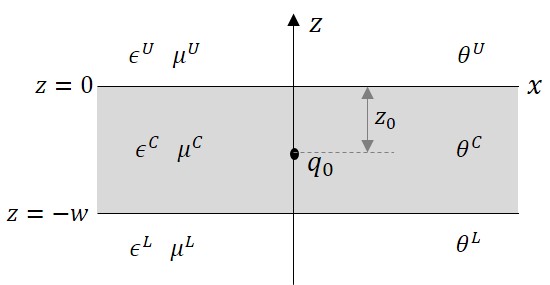}}
\caption{Schematics of the studied system and related parameters for the particular case of a point charge $q_0$ placed within the central (C) slab of width $w$ embedded between two semi-infinite media (upper, $U$, and lower, $L$, regions). The combinations of the $\theta$ parameters under study comprise the cases $(\theta^L, \theta^C, \theta^L) = (0, \pi, 0)$ (OI-TI-OI) and $(\pi, 0, \pi)$ (TI-OI-TI).}
\label{figesquema1}
\end{center}
\end{figure}

Recently, Y\=\i ng and Z{\"u}licke\cite{Yng-zulicke} used this method to derive analytical expressions for the magnetoelectric fields and probe the magnetoelectric response induced by a point charge on a thin, quasi-two dimensional slab of magnetoelectric material in a vacuum. 
Thin slabs are convenient models to study the TME in suspended layers of two-dimensional (graphene-like) materials \cite{Li11,devarakonda17} but also in a richer variety of systems, such as thin films \cite{okada16,pournaghavi21} or topological quantum wells \cite{bernevig06,Qi11}, where the material properties of the surrounding medium can make a difference. For this reason, in the present work we reformulate and extend the image dyon formalism to describe the magnetoelectric response in a more general case. The system is sketched in Figure \ref{figesquema1}: a point charge ($q_0$) is placed at an arbitrary location of the system composed by a thin slab of material $C$, sandwiched between an upper material ($U$) and a lower material ($L$), each with its own dielectric permittivity and magnetic permeability. The slab can be a TI ($\theta^C=\pi$) and the embedding materials OI ($\theta^L=\theta^U=0$), or viceversa.
Thus, general formulae for the dyons involved are provided for different combinations of the OI-TI planar interfaces configuring the slab, which are versatile enough to account for asymmetries in the parameters of the embedding materials, as well as for all possible combinations of external magnetization orientations at the two interfaces.
As in the image charge method \cite{jackson,kumagai,takagahara}, the analytical nature of the expressions gives intuitive insight into the influence of material parameters and slab dimensions on the resulting magnetoelectric fields. The equations are implemented in an accessible Mathematica code,\cite{mathcode} and are used to provide an overview on the formation of magnetic fields in thin slabs induced by a point charge.

\section{The method of image dyons}
\label{secImDy}

The method of image dyons can be thought as an extension of the method of image charges, broadly employed in electrostatics to account for dielectric discontinuities between adjacent materials in the determination of the field generated by a nearby electric point charge (see e.g. Refs. \cite{jackson,kumagai,takagahara}). 
The method consists in simulating the electrostatic potential $V$ of a dielectrically inhomogeneous system as the sum of the contributions of the real charge plus a set of virtual (\emph{image}) charges in a homogeneous dielectric. The image charges are considered as additional sources of electrostatic potential that contribute to shape the actual electrostatic field in the system, with the particularity that each of these contributions is restricted to the region of the space opposite to the location of the corresponding image charge with respect to the surface defining the dielectric discontinuity. 

In the case we are dealing with, however, the presence of interfaces with magnetoelectrics requires the additional consideration of (image) magnetic monopoles accompanying each of the electrostatic image charges to account for the magnetic fields arisen in the system. This renders new image objects called dyons $d_k=(q_k,g_k)$, where $q_k$ represents an image charge and $g_k$ a magnetic monopole, both located at the same position ${\mathbf r}_k$.

In systems with no time-dependent fields, Eqs. (\ref{eqFaraday}) and (\ref{eqAmpere}) turn into $\nabla \times {\mathbf E} = 0$ and $\nabla \times {\mathbf H} = 0$. Then, ${\mathbf E}$ and ${\mathbf B}$ can be obtained from the gradient of their scalar potentials, ${\mathbf E} = -\nabla V$ and ${\mathbf B} = -\nabla U$.\cite{jp22,arxivjp} In such conditions, and paralleling the method of image charges, the method of image dyons describes these potentials as

\begin{eqnarray}
\label{eqVpot}
V({\mathbf r}) &=& \frac{1}{\epsilon} \left[ \frac{q_0}{\left|{\mathbf r}-{\mathbf r}_0 \right|} + \sum_k \frac{q_k}{\left|{\mathbf r}-{\mathbf r}_k \right|} \right],\\
\label{eqUpot}
U({\mathbf r}) &=& \mu \sum_k \frac{g_k}{\left|{\mathbf r}-{\mathbf r}_k \right|},
\end{eqnarray}

\noindent where, for convenience, the dielectric constant $\epsilon$ and the magnetic permeability $\mu$ are chosen as those of the region where the source charge $q_0$ is located, and $k$ runs over all dyons needed to define the potentials. Then, $q_k$ and $g_k$ values are yielded by applying boundary conditions (BCs) at the interfaces which account for these \emph{Ans{\"a}tze}. 
In the presence of a single TI-OI interface, two image dyons suffice to fulfill the boundary conditions.\cite{Qi11,arxivjp} By contrast, in a finite slab the double interface makes the summations in Eqs. (\ref{eqVpot}) and (\ref{eqUpot}) constitute infinite series. We elaborate on this point below, as a proper derivation of the dyon series implies a few conceptual subtleties.

\begin{figure*}
\begin{center}
\resizebox{0.95\textwidth}{!}{\includegraphics{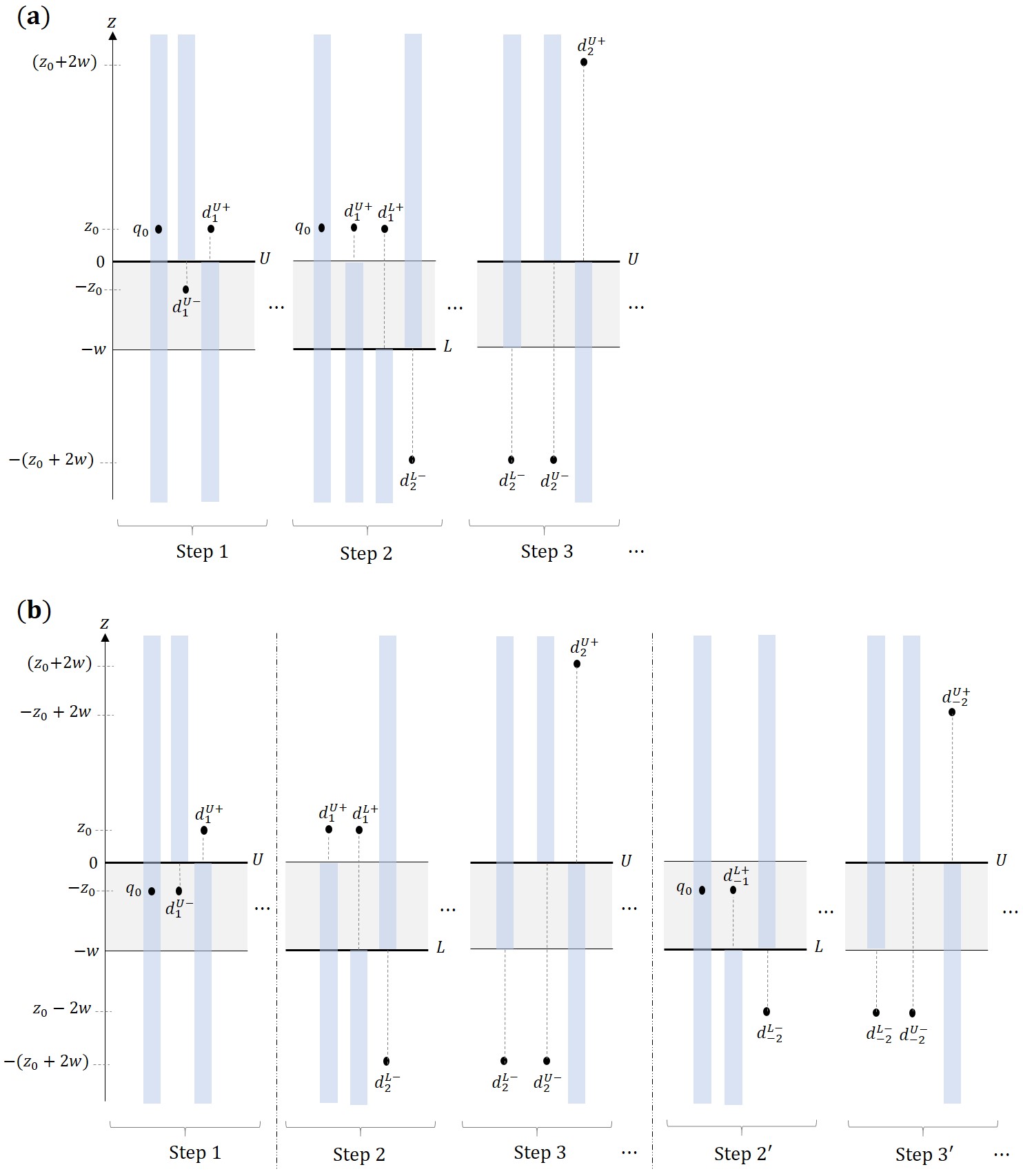}}
\caption{Schematic of the first iterations of the image dyon method for a source charge $q_0$ placed near (a) and inside (b) the QW slab. The blue-shaded region vertically aligned with each charge/dyon represents the region where its contribution to the potentials has to be considered. The dyons have been offset in the horizontal axis for clarity, but they all share the same position in the $xy$ plane. The interface where the boundary conditions are intended to be satified in each step is highlighted by a thicker line. In (a), the iteration process alternates continuously between the two interfaces $U$ and $L$, generating an infinite set of image dyons. In contrast, the process in (b) splits after the first step into two (infinite) $d_k^{m\nu}$ subsets: one labelled with $k>0$ and the other with $k<0$ subscripts. Note that $d_{\pm 1}^{L-}$ and $d_{-1}^{U \nu}$ do not exist by construction.} 
\label{figesquema2}
\end{center}
\end{figure*}

Figure \ref{figesquema2}(a) depicts the first few steps of the process carried out to obtain the general expressions of the (infinite) series of image dyons when the source charge is placed in the upper region, near the slab. Here we follow the scheme and nomenclature proposed by Y\=\i ng and Z{\"u}licke.\cite{Yng-zulicke} Thus, dyons are expressed as $d_k^{m\nu}$, where the supersctipt $m=U$ ($L$) denotes dyons obtained by mirroring in the upper (lower) interface, and $\nu = +$ ($-$) when the location of the image dyon is above (below) $m$. Additionally, the subscript $k=\pm 1, \pm 2, \dots$ is related to the distance between the image and a reference interface. It will become clear from the figure that the position of a dyon is given by:

\begin{equation}
\label{eqLoc}
{\mathbf r}_k^{\nu}=(0,0,\nu \text{sgn}(k)\{z_0+2[k-\text{sgn}(k)]w\}),
\end{equation}

\noindent where $z_0$ is the distance of the source (real) charge to the $xy$ plane and $w$ is the width of the slab.

To start the process (Step 1 in Figure \ref{figesquema2}(a)), we apply the boundary conditions derived from the constitutive relations (\ref{eqConstD}) and (\ref{eqConstH}) to the upper interface only, while ignoring the presence of the lower interface. 
That is,  ${\bf D}_{\perp}^U = {\bf D}_{\perp}^C$, 
${\bf H}_{\parallel}^U = {\bf H}_{\parallel}^C$, ${\bf E}_{\parallel}^U = {\bf E}_{\parallel}^C$, ${\bf B}_{\perp}^U = {\bf B}_{\perp}^C$. As in the image charge method, it is convenient to place the virtual dyons as mirror images of the source (i.e., equidistant with respect to the interface). 
We then obtain the first set of image dyons, i.e., $d_1^{U+}$ at $z=z_0$ and $d_1^{U-}$ at $z=-z_0$, whose contributions to $V$ and $U$ (in $z<0$ and $z>0$, respectively) ensure the proper continuity of the fields across the interface.

For the sake of clarity, the regions of $z$ where each dyon contributes to $V$ and $U$ have been shaded in the figure. 
For instance, $d_1^{U-}$ must be considered only when describing the potentials in the region $U$.
Two consequences follow: (i) it becomes clear that the summations in Eqs. (\ref{eqVpot}) and (\ref{eqUpot}) will run over specific sets of dyons, depending on the region where $V$ and $U$ are calculated; and (ii) $d_1^{U+}$ must be taken into account in the next step, which consists in ensuring the fulfilment of the boundary conditions at the lower interface. As shown in Fig. \ref{figesquema2}(a) (see Step 2), this is accomplished by incorporating a new set of dyons, $d_1^{L+}$ and $d_2^{L-}$, equidistant from the $L$ interface, and considering both, $q_0$ and $d_1^{U+}$ as the source charges. However, the dyon $d_2^{L-}$ unbalances the boundary conditions in the upper interface, so that new dyons, $d_2^{U+}$ and $d_2^{U-}$, will be needed in order to fix it (Step 3), which, in turn, will provoke further imbalances of the BCs in the lower interface. The subsequent rearrangements of the potentials and related dyons involved let us foresee the infinite character of the method and of the summations in Eqs. (\ref{eqVpot}) and (\ref{eqUpot}). Fortunately, the convergence of the method is fast enough as not to require more than a few steps to reach considerable accuracy.

When the source charge is placed inside the slab the procedure becomes somewhat more involved, since now two differentiated, infinite subsets of image dyons appear. This can be seen in Fig. \ref{figesquema2}(b). The first step is similar to the case with the source charge outside the slab, and implies the obtainment of the dyons $d_1^{U+}$ at $z=z_0$ and $d_1^{U-}$ at $z=-z_0$. Hereafter, however, the image mirroring process follows two different paths, each considering $d_1^{U+}$ or $q_0$ as sources, thereby generating the subset of dyons labelled with $k>0$ and $k<0$ subscripts, respectively (see the Figure). The origin of this separation is the different location of the afore-mentioned source charge and dyon.

Following the process described above, one obtains the location of the dyons contributing to $V$ and $U$ in Eqs. (\ref{eqVpot}) and (\ref{eqUpot}). 
The magnitude of the charge and magnetic monopole follow from the imposition of the boundary conditions on the due interfaces. Because these are individual interfaces, the procedure to this end is the same as described elsewhere \cite{arxivjp}. 
Below we provide the general expressions for a point charge near or inside a slab.
It is worth stressing that the sets of equations are independent of the particular configuration of the system, i.e., of the topological/trivial insulator nature of the slab material. Thus, they apply to both, a topologic slab surrounded by ordinary media and an ordinary (e.g., semiconductor) slab between topological insulators.

\subsection{Image dyons and potentials from a point charge placed near the central slab}

Without loss of generality, we assume the source charge $q_0$ to be located at ${\mathbf r}_0=(0,0,z_0)$, with $z_0 > 0$ (upper medium).
In such a case, the corresponding image dyons $d_k^{m\nu} = (q_k^{m\nu}, g_k^{m\nu})$ are obtained by means of the following recurrence relations:

\begin{equation}
\label{eqMqU1}
\left(\begin{matrix} q_k^{U\nu}\\  g_k^{U\nu} \end{matrix}\right)=\beta^U \left(\begin{matrix} -\xi^U & -\eta^U (\frac{2 \epsilon^U \mu^U}{\mu^C}) \alpha\\ -\nu \eta^U (\frac{2 \epsilon^C}{\epsilon^U \mu^U}) \alpha & \nu \chi^U \end{matrix}\right) \left(\begin{matrix} q_k^{L-}\\  g_k^{L-} \end{matrix}\right),
\end{equation}

\begin{equation}
\label{eqMqU2}
\left(\begin{matrix} q_k^{L+}\\  g_k^{L+} \end{matrix}\right)=\beta^L \left(\begin{matrix} -\xi^L & \eta^L (\frac{2 \epsilon^U \mu^U}{\mu^C}) \alpha\\ -\eta^L (\frac{2 \epsilon^C}{\epsilon^U \mu^U}) \alpha & -\chi^L \end{matrix}\right) \left(\begin{matrix} q_k^{U+}\\  g_k^{U+} \end{matrix}\right),
\end{equation}

\begin{equation}
\label{eqMqU3}
\left(\begin{matrix} q_k^{L-}\\  g_k^{L-} \end{matrix}\right)= \left(\begin{matrix} 1 & 0\\ 0 & -1 \end{matrix}\right) \left(\begin{matrix} q_{k-1}^{L+}\\  g_{k-1}^{L+} \end{matrix}\right), 
\end{equation}

\noindent with the following initial values:

\begin{equation}
\label{eqMqU01}
\left(\begin{matrix} q_1^{U\nu}\\  g_1^{U\nu} \end{matrix}\right)=\beta^U \left(\begin{matrix} \xi^U - 2 \alpha^2 & 0\\ -\nu \eta^U (\frac{2}{\mu^U}) \alpha & 0 \end{matrix}\right) \left(\begin{matrix} q_0\\ 0 \end{matrix}\right),
\end{equation}

\begin{equation}
\label{eqMqU02}
\left(\begin{matrix} q_1^{L+}\\  g_1^{L+} \end{matrix}\right)=\beta^L \left(\begin{matrix} -\xi^L & \eta^L (\frac{2 \epsilon^U \mu^U}{\mu^C}) \alpha\\ -\eta^L (\frac{2 \epsilon^C}{\epsilon^U \mu^U}) \alpha & -\chi^L \end{matrix}\right) \left(\begin{matrix} q_0 + q_1^{U+}\\  g_1^{U+} \end{matrix}\right). 
\end{equation}

Notice that $d_1^{L-}$ does not exist by construction.

In the above equations, we have defined $\beta^m = [(\epsilon^m+\epsilon^C)(1/\mu^m+1/\mu^C)+\alpha^2]^{-1}$, $\xi^m = (\epsilon^m-\epsilon^C)(1/\mu^m+1/\mu^C)+\alpha^2$, $\chi^m = (\epsilon^m+\epsilon^C)(1/\mu^m-1/\mu^C)+\alpha^2$, and $\eta^m = \text{sgn} [{\mathbf k} \cdot {\mathbf M}^m]$, which depends on the orientation with respect to the vector ${\mathbf k}$ (the unitary vector in the positive direction of the $z$ axis) of the external magnetization ${\mathbf M}^m$ lifting the time reversal symmetry near the corresponding ($m=U,L$) interface.\cite{Qi09,Qi11,Campos,sekine} Note that $\eta$ should not be confused with $P$ in Eqs. (\ref{eqConstD})-(\ref{eqConstH}). Although they are very closely related (both determine the sign of the magnetoelectric terms), they are referred to different orientation vectors. Here we used $\eta$ in order to obtain more general expressions.

The \emph{Ans{\"a}tze} employed for the electric $V({\mathbf r})$ and magnetic $U({\mathbf r})$ potentials, Eqs. (\ref{eqVpot}) and (\ref{eqUpot}), can be expanded in terms of the above image dyons as follows:

\begin{eqnarray}
\label{eqPOTSqU1}
V^{UU}({\mathbf r})&=&\frac{1}{\epsilon^U}\left[\frac{q_0}{\left|{\mathbf r}-{\mathbf r}_0 \right|}+\sum_{m=U,L}\sum^{\infty}_{k=1}\frac{q_k^{m-}}{\left|{\mathbf r}-{\mathbf r}_k^-  \right|}\right],\\
\label{eqPOTSqU2}
V^{UC}({\mathbf r})&=&
\begin{aligned}[t]
&\frac{1}{\epsilon^U}\frac{q_0}{\left|{\mathbf r}-{\mathbf r}_0 \right|} \\
&+\frac{1}{\epsilon^U}\sum^{\infty}_{k=1} \left( \frac{q_k^{U+}}{\left|{\mathbf r}-{\mathbf r}_k^+  \right|} +\frac{q_k^{L-}}{\left|{\mathbf r}-{\mathbf r}_k^-  \right|} \right),
\end{aligned} \\
\label{eqPOTSqU3}
V^{UL}({\mathbf r})&=&\frac{1}{\epsilon^U}\left[\frac{q_0}{\left|{\mathbf r}-{\mathbf r}_0 \right|}+\sum_{m=U,L}\sum^{\infty}_{k=1}\frac{q_k^{m+}}{\left|{\mathbf r}-{\mathbf r}_k^+  \right|}\right],\\
\label{eqPOTSqU4}
U^{UU}({\mathbf r})&=&\mu^U \sum_{m=U,L} \sum^{\infty}_{k=1} \frac{g_k^{m-}}{\left|{\mathbf r}-{\mathbf r}_k^- \right|},\\
\label{eqPOTSqU5}
U^{UC}({\mathbf r})&=&\mu^U \sum^{\infty}_{k=1} \left( \frac{g_k^{U+}}{\left|{\mathbf r}-{\mathbf r}_k^+  \right|} +\frac{g_k^{L-}}{\left|{\mathbf r}-{\mathbf r}_k^-  \right|} \right),\\
\label{eqPOTSqU6}
U^{UL}({\mathbf r})&=&\mu^U \sum_{m=U,L} \sum^{\infty}_{k=1} \frac{g_k^{m+}}{\left|{\mathbf r}-{\mathbf r}_k^+ \right|}.
\end{eqnarray}

\noindent where the first superscrit in $V$ and $U$ indicates the region in which the source charge is located, and the second the region in which the potential is described. As explained previously, in Eqs. (\ref{eqPOTSqU1})-(\ref{eqPOTSqU6}) only the dyons accounting for the fields in each region are considered.

\subsection{Image dyons and potentials from a point charge placed within the slab}

Now we consider $q_0$ placed at ${\mathbf r}_0=(0,0,-z_0)$, with $w > z_0 > 0$. In this case, the recurrence relations for the subset of image dyons with $k>0$ (see section \ref{secImDy}) read:

\begin{equation}
\label{eqMqC1}
\left(\begin{matrix} q_k^{U\nu}\\  g_k^{U\nu} \end{matrix}\right)=\beta^U \left(\begin{matrix} -\xi^U & -\eta^U (2 \epsilon^C) \alpha\\ -\nu \eta^U (\frac{2}{\mu^C}) \alpha & \nu \chi^U \end{matrix}\right) \left(\begin{matrix} q_k^{L-}\\  g_k^{L-} \end{matrix}\right),
\end{equation}

\begin{equation}
\label{eqMqC2}
\left(\begin{matrix} q_k^{L+}\\  g_k^{L+} \end{matrix}\right)=\beta^L \left(\begin{matrix} -\xi^L & \eta^L (2 \epsilon^C) \alpha\\ -\eta^L (\frac{2}{\mu^C}) \alpha & -\chi^L \end{matrix}\right) \left(\begin{matrix} q_k^{U+}\\  g_k^{U+} \end{matrix}\right), 
\end{equation}

\begin{equation}
\label{eqMqC3}
\left(\begin{matrix} q_k^{L-}\\  g_k^{L-} \end{matrix}\right)= \left(\begin{matrix} 1 & 0\\ 0 & -1 \end{matrix}\right) \left(\begin{matrix} q_{k-1}^{L+}\\  g_{k-1}^{L+} \end{matrix}\right), 
\end{equation}

\noindent with the following initial values:

\begin{equation}
\label{eqMqC01}
\left(\begin{matrix} q_1^{U\nu}\\  g_1^{U\nu} \end{matrix}\right)=\beta^U \left(\begin{matrix} -\xi^U & 0\\ -\nu \eta^U (\frac{2}{\mu^C}) \alpha & 0 \end{matrix}\right) \left(\begin{matrix} q_0\\ 0 \end{matrix}\right).
\end{equation}

For the subset of image dyons with $k<0$, the recurrence relations are given by equations (\ref{eqMqC1}) and (\ref{eqMqC2}), together with 

\begin{equation}
\label{eqMqC4}
\left(\begin{matrix} q_k^{L-}\\  g_k^{L-} \end{matrix}\right)= \left(\begin{matrix} 1 & 0\\ 0 & -1 \end{matrix}\right) \left(\begin{matrix} q_{k+1}^{L+}\\  g_{k+1}^{L+} \end{matrix}\right)
\end{equation}

\noindent and the initial values:

\begin{equation}
\label{eqMqC02}
\left(\begin{matrix} q_{-1}^{L+}\\  g_{-1}^{L+} \end{matrix}\right)=\beta^L \left(\begin{matrix} -\xi^L & 0\\ -\eta^L (\frac{2}{\mu^C}) \alpha & 0 \end{matrix}\right) \left(\begin{matrix} q_0\\ 0 \end{matrix}\right).
\end{equation}

Note that in this case ($q_0$ in the central slab), $d_{\pm 1}^{L-}$ and $d_{-1}^{U\nu}$ do not exist by construction.

The scalar potentials in terms of these image charges are written as

\begin{eqnarray}
\label{eqPOTSqC1}
V^{CU}({\mathbf r})&=&\frac{1}{\epsilon^C}\left[\frac{q_0}{\left|{\mathbf r}-{\mathbf r}_0 \right|}+\sum_{m=U,L}\sum^{\infty}_{k=1}\frac{q_{\pm k}^{m-}}{\left|{\mathbf r}-{\mathbf r}_{\pm k}^-  \right|}\right],\\
\label{eqPOTSqC2}
V^{CC}({\mathbf r})&=&
\begin{aligned}[t]
&\frac{1}{\epsilon^C}\frac{q_0}{\left|{\mathbf r}-{\mathbf r}_0 \right|} \\
&+\frac{1}{\epsilon^C}\sum^{\infty}_{k=1} \left( \frac{q_{\pm k}^{U+}}{\left|{\mathbf r}-{\mathbf r}_{\pm k}^+  \right|} +\frac{q_{\pm k}^{L-}}{\left|{\mathbf r}-{\mathbf r}_{\pm k}^-  \right|} \right),
\end{aligned} \\
\label{eqPOTSqC3}
V^{CL}({\mathbf r})&=&\frac{1}{\epsilon^C}\left[\frac{q_0}{\left|{\mathbf r}-{\mathbf r}_0 \right|}+\sum_{m=U,L}\sum^{\infty}_{k=1}\frac{q_{\pm k}^{m+}}{\left|{\mathbf r}-{\mathbf r}_{\pm k}^+  \right|}\right],\\
\label{eqPOTSqC4}
U^{CU}({\mathbf r})&=&\mu^C \sum_{m=U,L} \sum^{\infty}_{k=1} \frac{g_{\pm k}^{m-}}{\left|{\mathbf r}-{\mathbf r}_{\pm k}^- \right|},\\
\label{eqPOTSqC5}
U^{CC}({\mathbf r})&=&\mu^C \sum^{\infty}_{k=1} \left( \frac{g_{\pm k}^{U+}}{\left|{\mathbf r}-{\mathbf r}_{\pm k}^+  \right|} +\frac{g_{\pm k}^{L-}}{\left|{\mathbf r}-{\mathbf r}_{\pm k}^-  \right|} \right),\\
\label{eqPOTSqC6}
U^{CL}({\mathbf r})&=&\mu^C \sum_{m=U,L} \sum^{\infty}_{k=1} \frac{g_{\pm k}^{m+}}{\left|{\mathbf r}-{\mathbf r}_{\pm k}^+ \right|}.
\end{eqnarray}

\section{Topological magnetoelectric field in a quantum well}

Building up on the previous equations, the dyons of a thin slab can be computed by means of a recurrent procedure. A Mathematica code implementing such a procedure can be found in Ref. \cite{mathcode}.

For illustrative purposes, in this section we use the code to study the behavior of magnetoelectric fields induced by a source charge on slabs of different materials and thicknesses.

\begin{figure}[htb]
\begin{center}
\includegraphics[width=1.0\columnwidth]{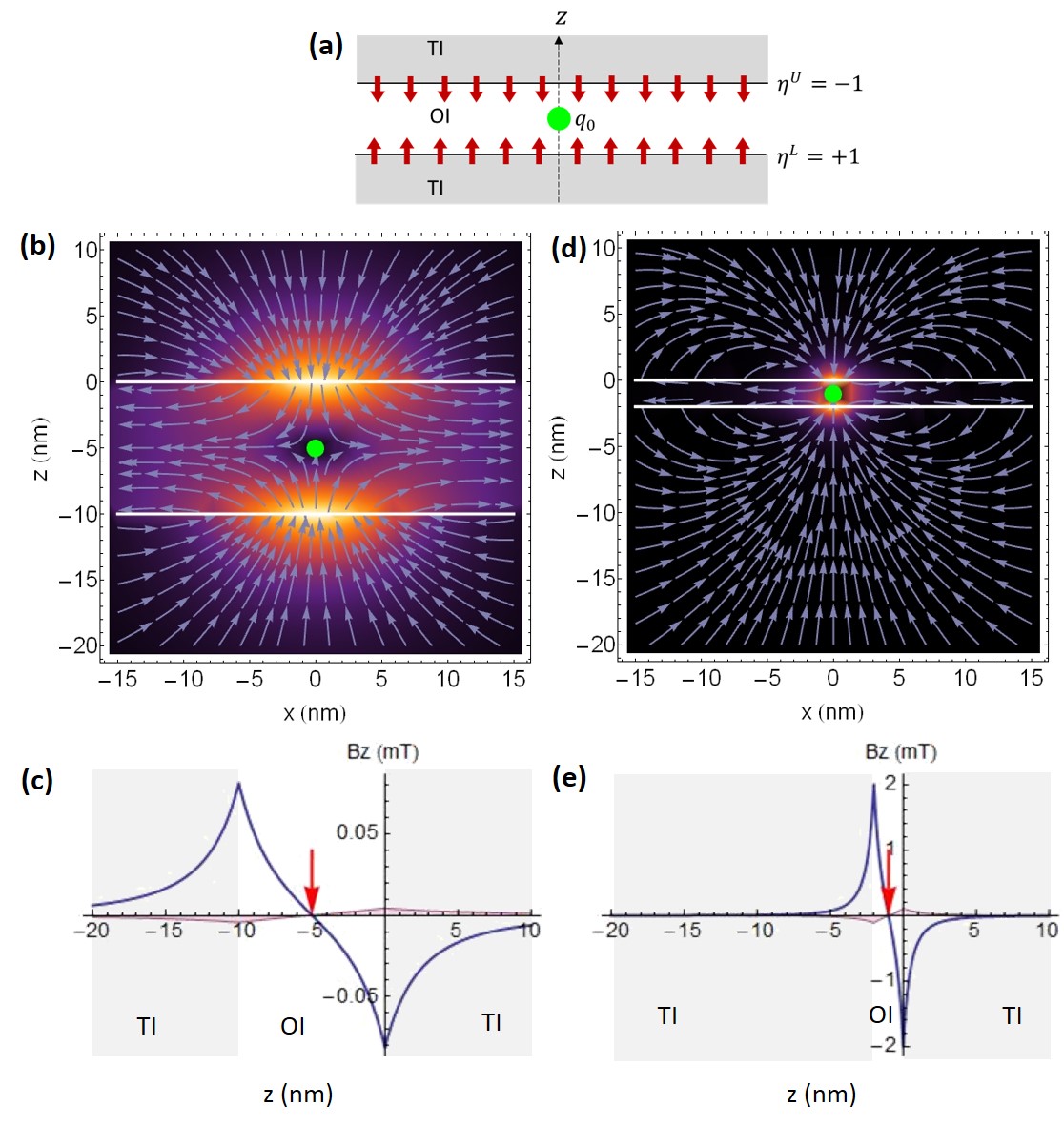}
\caption{Magnetic field induced by a charge $q_0=+1$ a.u. at the center of an OI slab within a TI (TI-OI-TI system) assuming antiparallel orientations of the external magnetization on the interfaces, represented by red arrows in (a). 
(b) [(d)] Stream density plots for a $w=10$ nm-thick [$w=2$ nm-thick] slab. The position of $q_0$ is denoted by the green dot, and the thick, white lines indicate the upper and lower interfaces of the slab. 
(c) [(e)] Corresponding $z$ projection of ${\bf B}$ along the vertical axis containing $q_0$, whose position is pointed out by the red arrow. TI materials are indicated by the shaded regions. 
In (c) [(e)] the pink line quantifies the difference between the $B_z$ profile for the TI-OI-TI system and that for its OI-TI-OI counterpart.}
\label{figq0cent}
\end{center}
\end{figure}

We start by setting representative parameters for the OI $(\epsilon_{\text{OI}}, \mu_{\text{OI}}, \theta_{\text{OI}})=(5, 1, 0)$ and the TI $(\epsilon_{\text{TI}}, \mu_{\text{TI}}, \theta_{\text{TI}})$=$(10, 1, \pi)$ materials in the system.
Two slab thicknesses are considered, namely $w=10$ nm and $w=2$ nm. These material parameters and dimensions are representative of thick and thin semiconductor quantum wells. We then choose the slab to be made of an OI and the surroundings of TI.
In Fig. \ref{figq0cent}, we study the magnetic field ${\bf B}$ arising from a point charge $q_0=+1$ a.u. in the center of the slab, assuming that the external magnetization points towards the slab, see Fig. \ref{figq0cent}(a). The figure shows stream density plots of the magnetic field, as well as the $z$-projection of the magnetic field along the $z$-axis (the axis containing the source charge). This magnitude can be taken as a paradigmatic cross section for the comparison of absolute values of ${\bf B}$, because it contains the maximum values of the induced field, as can be inferred from the stream density plots.

One can see in Figure \ref{figq0cent} that ${\bf B}$ is suppressed around $q_0$ and a nodal plane for $B_z$ builds up at that location (Figs. \ref{figq0cent}(b) and \ref{figq0cent}(c)). The same occurs for narrower slabs (Figs. \ref{figq0cent}(d) and (e)). This result is in line with the trends reported in recent works \cite{jp22,Yng-zulicke}, and can be rationalized in terms of the sense of rotation of the Hall currents triggered by the point charge on the $U$ and $L$ interfaces.\cite{jp22} It should be noted, however, that our calculations do not show the vortexlike pattern found in Ref. \cite{Yng-zulicke}, for any set of material parameters (not shown). 
Regarding the effect of the slab thickness, Fig. \ref{figq0cent} shows that narrowing down the slab increases the magnetic field strength because the interfaces are closer to the source charge (cf. $B_z$ in panels (c)  and (e)), and reorients the field (cf. stream density plots, panels (b) and (d)).

\begin{figure}[htb]
\begin{center}
\resizebox{1.0\columnwidth}{!}{\includegraphics{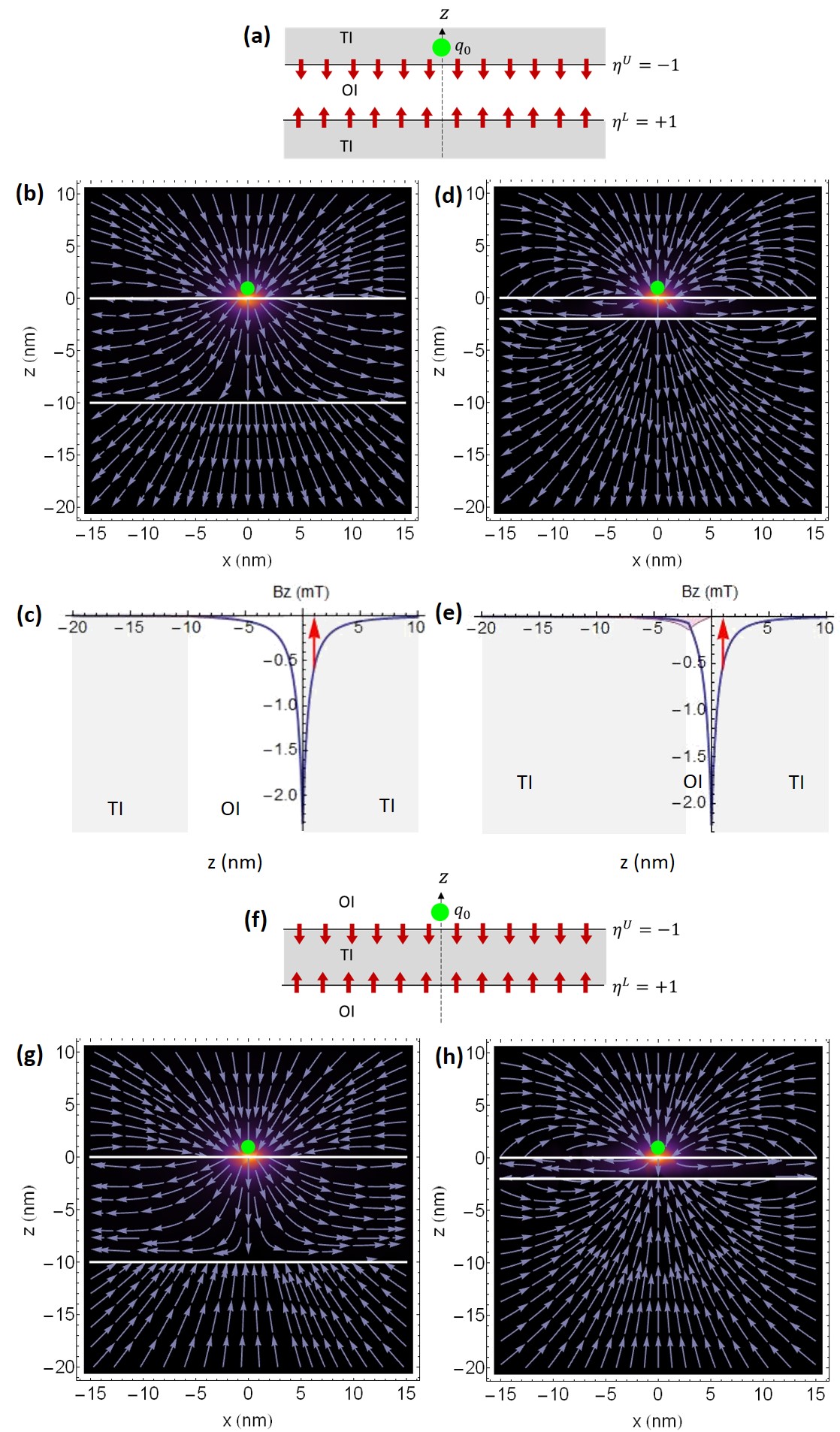}}
\caption{(a)-(e) Same as in Fig. \ref{figq0cent} but for $q_0$ placed at $z_0=1$ nm above the upper interface. (g) [(h)] Same as in (b) [(d)] but for the OI-TI-OI system (f).}
\label{figq0out}
\end{center}
\end{figure}

When the slab is made of a TI material instead (OI-TI-OI configuration), the resulting magnetic fields are similar to those described above. The only differences are ascribable to the different distribution of the dielectric constants in the system. For the constants we consider, this leads to |${\bf B}$| being slightly smaller. 
This can be inferred from the pink lines in Fig. \ref{figq0cent}(c) and Fig. \ref{figq0cent}(e), which represent the difference between $B_z$ for a TI slab and that for an OI slab.

Figure \ref{figq0out}(a-e) is analogous to Fig. \ref{figq0cent}, but now the source charge has been moved at a distance $z_0=1$ nm on top of the upper interface, see panel (a). Magnetic fields build up near the TI-OI interface in the vicinity of the source charge, which are fairly unsensitive to the thickness of the slab.\cite{jp22} 

Interestingly, for this charge location, switching from an OI slab to a TI one brings about qualitative differences. 
Namely, the stream density plots (Figs. \ref{figq0out}(g) and \ref{figq0out}(h)) show opposite orientations of the magnetic field below the lower interface, as compared to the OI slab case (Figs. \ref{figq0out}(b) and \ref{figq0out}(d)). 
This is a signature of the relevance on the magnetoelectric coupling of the polarization charges induced at the interfaces as a result of the different dielectric response of adjacent materials. Thus, in the upper interface, near the electrostatic source charge, the magnetoelectric coupling is dominated by the direct Coulomb term, yielding a qualitatively similar magnetic field distribution. However, in the lower interface, further from the source charge, the effect of the surface polarization charge prevails. Such polarization charge presents in the OI-TI-OI system opposite sign than in TI-OI-TI, thus explaining the opposite orientation of the induced magnetic field.
 
\begin{figure}[htb]
\begin{center}
\resizebox{1.0\columnwidth}{!}{\includegraphics{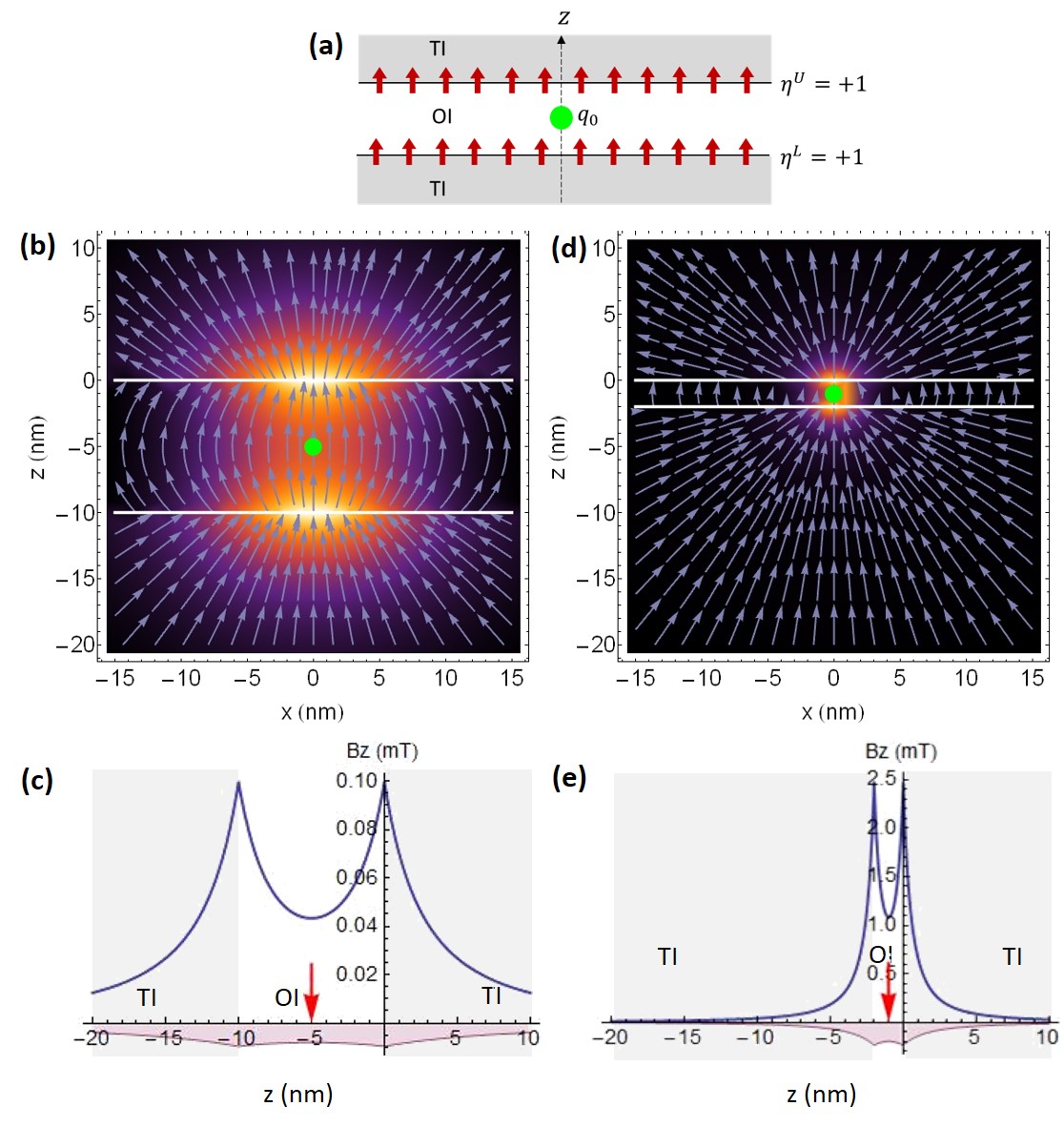}}
\caption{Same as in Fig. \ref{figq0cent} but for the same orientation of the external magnetizations at both interfaces.}
\label{figq0centMparallel}
\end{center}
\end{figure}

In the examples considered so far, magnetization on the two interfaces was antiparallel, i.e., $\text{sign}[{\bf M}^U \cdot {\bf n}^U] = \text{sign}[{\bf M}^L \cdot {\bf n}^L]$. 
In particular, ${\bf M}^m$ pointed inwards the TI.
However, our formalism is general enough as to also describe
(i) the effect of parallel magnetization on the interfaces, and 
(ii) the asymmetry caused by different $U$ and $L$ materials. 
Figure \ref{figq0centMparallel} presents results obtained assuming (i). They are directly comparable with those in Fig. \ref{figq0cent}, for $\eta^U=-1 \rightarrow \eta^U=+1$ is the only change between them.
The comparison evidences that the magnetic field distribution presents now a preferential orientation dictated by that of ${\bf M}$, eliminating thus the $B_z$ nodal plane and the vortexlike patterns around the interfaces. 
This is because the Hall currents induced on both interfaces have now the same sense of rotation, and hence reinforce $|{\bf B}|$ throughout the system. 
This reinforcement entails larger differences between the OI-TI-OI and the TI-OI-TI systems (see pink lines in Figs. \ref{figq0centMparallel}(c) and \ref{figq0centMparallel}(e)), ascribable once again to the different spatial distribution of the dielectric response.

The most remarkable feature of the field distribution is its resemblance with that of a Pearl vortex, previously reported for a single OI-TI interface between two semi-infinite media.\cite{jp22,Nogueira2022}
Indeed, the slab can be considered as a transition region between those presenting the characteristic distribution of a Pearl vortex, with the difference that now its prominent features (as e.g., the straight stream lines) show up in the same type of material (TI in Fig. \ref{figq0centMparallel}(b)).

\begin{figure}[htb]
\begin{center}
\resizebox{1.0\columnwidth}{!}{\includegraphics{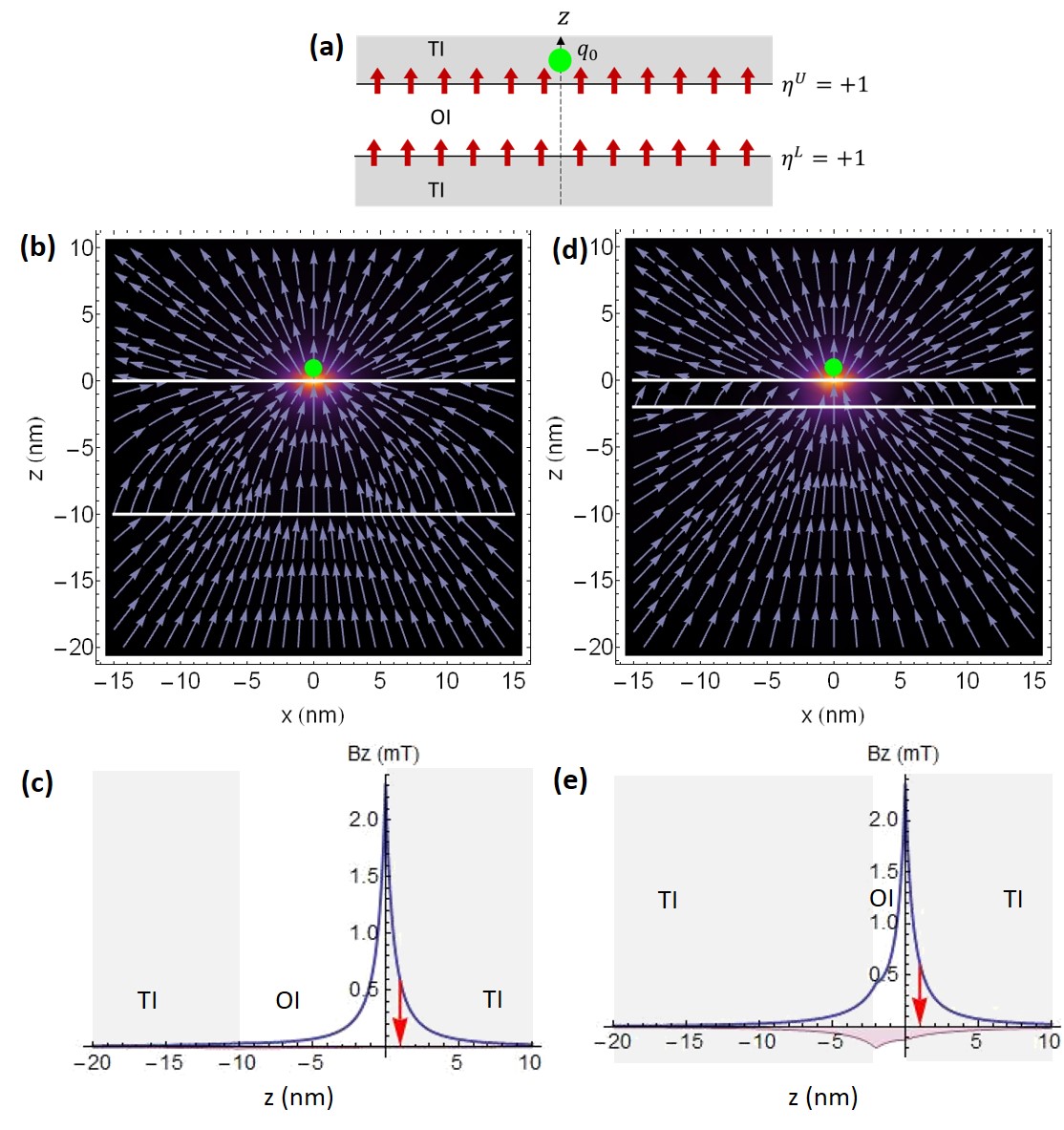}}
\caption{Same as in Fig. \ref{figq0out}(a)-(e) but for the same orientation of the external magnetizations at both interfaces.}
\label{figq0outMparallel}
\end{center}
\end{figure}

The Pearl vortex-like distribution is distorted when asymmetries come into play. A representative example can be seen in Fig. \ref{figq0outMparallel}(b), similar to those in Fig. \ref{figq0centMparallel} but with $q_0$ moved into the upper medium. The distribution of the magnetic field in the vicinity of the $U$ interface evidences a deviation from the ideal profile of a Pearl vortex. Indeed, one can observe a slight curvature of the $z>0$ stream lines near the interface. This effect also occurs when the off-centered source charge remains within the central slab (not shown). However, in both cases the resemblance of the field distribution to a Pearl vortex is recovered for thin enough slabs (see Fig. \ref{figq0outMparallel}(d)).

\begin{figure}[t]
\begin{center}
\resizebox{1.0\columnwidth}{!}{\includegraphics{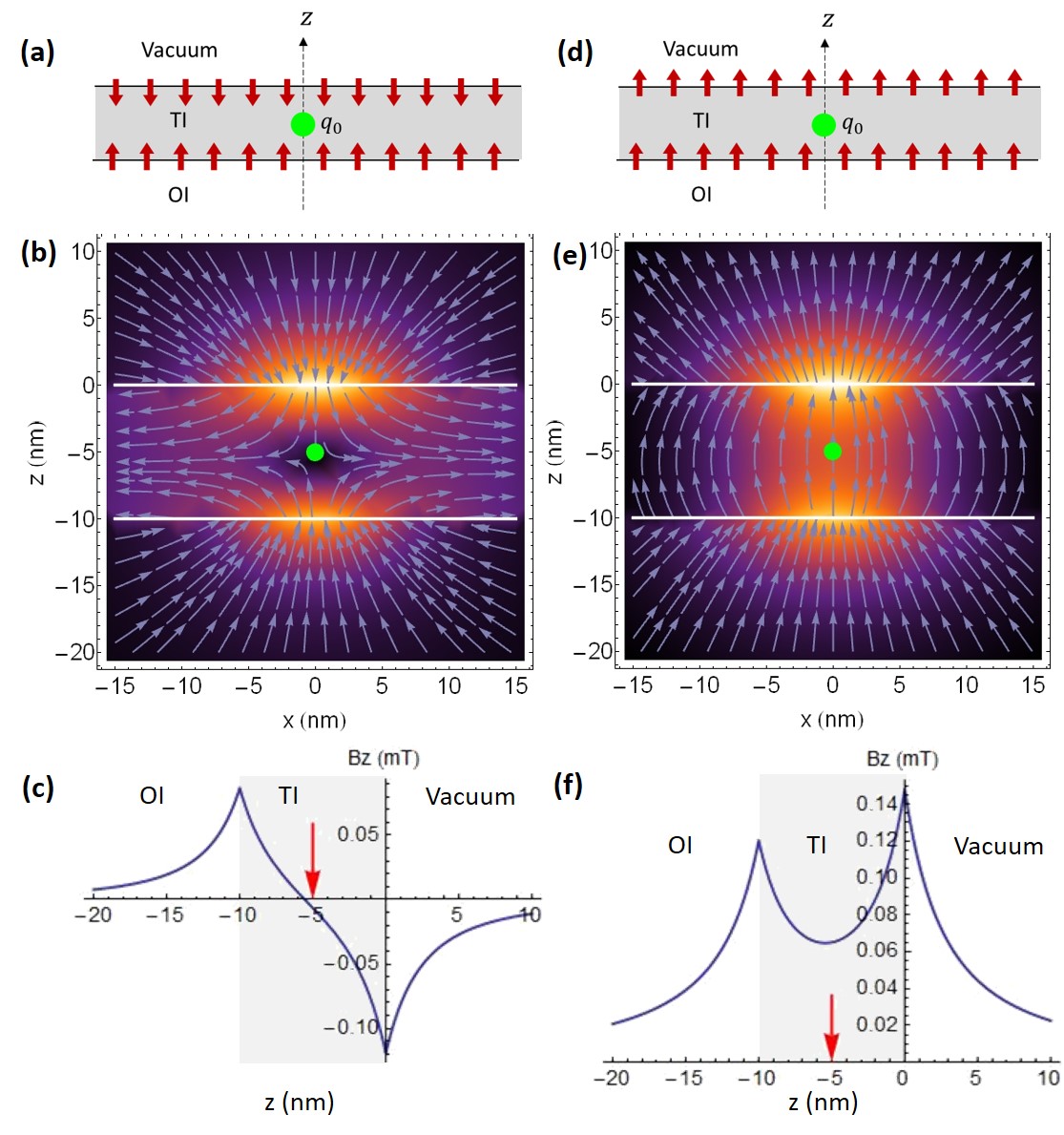}}
\caption{(a)-(c) [(d)-(f)] Same as in Fig. \ref{figq0cent}(a)-(c) [Fig. \ref{figq0centMparallel}(a)-(c)] but for a OI-TI-OI slab with $(\epsilon^U, \mu^U, \theta^U) = (1, 1, 0)$ (i.e., vacuum) and $(\epsilon^L, \mu^L, \theta^L) = (5, 1, 0)$ (i.e., semiconductor).}
\label{figq0centVacuum}
\end{center}
\end{figure}

In actual quantum wells, an additional source of asymmetry may stem from having different $U$ and $L$ media. 
This configuration (particularly, a TI of finite thickness with two surfaces, with vacuum in one side and a semiconductor substrate on the other) was proposed as a possible setup to observe the topological magnetoelectric effect through measurements of Faraday and Kerr rotations induced on the TR symmetry-breaking surfaces,\cite{maciejko10,tse10} and was later on experimentally realized confirming the expected quantization of the magnetoelectric coupling in TIs.\cite{wu16,dziom17} 
This scenario is studied in Fig. \ref{figq0centVacuum}, where two different ordinary media given by $(\epsilon^U, \mu^U, \theta^U) = (1, 1, 0)$ (i.e., vacuum), and $(\epsilon^L, \mu^L, \theta^L) = (5, 1, 0)$ (e.g., a semiconductor), delimit the boundaries of a TI slab with $(\epsilon^C, \mu^C, \theta^C) = (10, 1, \pi)$. 
All calculations in the figure keep $q_0$ at the center of the slab. 
A number of visible differences with respect to the case of symmetric material distribution appear. 
The plots reveal stronger magnetic fields in the upper interface region, 
the $B_z$ nodal plane disappears in the case of antiparallel magnetizations (see Fig. \ref{figq0centVacuum}(b) away from $x=0$), and the Pearl vortex pattern becomes slightly distorted for $z>0$ in the case of parallel magnetizations (see Fig. \ref{figq0centVacuum}(e)).
In addition, the strength of the magnetic field near the $U$ interface increases as compared with Figs. \ref{figq0cent} and \ref{figq0centMparallel}, which is a consequence of the larger polarization charge induced in the TI-vacuum interface due to their larger dielectric contrast.

\section{Concluding remarks}

We have introduced a compact implementation of the image dyons method which provides the magnetoelectric fields induced by a point charge near or inside a quasi-two dimensional slab with OI-TI interfaces. The implementation is general enough as to account for all possible combinations of (i) the OI-TI planar interfaces configuring the slab, (ii) the material parameters, and (iii) the orientations of the local magnetizations needed to lift the surface time-reversal symmetry on each interface.
The equations can be readily implemented in a recurrent calculation routine, like that we provide in Ref. \cite{mathcode}, which ensures speed, accuracy and a fast convergence of the derived electric and magnetic fields.
The illustrative calculations highlight the rich variety of the magnetoelectric response depending on the system configuration, which is certainly an incentive to custom-design the resulting magnetic field.

\begin{acknowledgments}
We acknowledge support from MICINN project PID2021-128659NB-I00, 
UJI project B-2021-06 and Generalitat Valenciana Prometeo project 22I235-CIPROM/2021/078.
\end{acknowledgments}

\end{document}